\documentstyle[preprint,aps]{revtex}
\begin{document}
\preprint{IMSc/94/17}
%
%
%
%

\title{Failure of Fermi Liquid Theory in 2-D :\\
 A Signal from Perturbation Theory}

\author{G. Baskaran\cite{email}}

\address{ The Institute of Mathematical Sciences,
                      Madras 600 113, India}

\maketitle

\begin{abstract}
We study the perturbative correction to the ground state energy
eigenvalue of a 2-dimensional dilute fermi gas with weak
short-range two body repulsion.  From
the structure of the energy shift we infer the presence of
an induced two body long range repulsive interaction
$ {\displaystyle {1\over{r^2}}
}$ among the constituent electrons
indicating a potential instability of the fermi liquid ground state.
\end{abstract}


The discovery of high $T_c$
superconductivity has led to the revival of interest in the physics of
strongly correlated electrons in two dimesions and the
possibility of the failure of fermi liquid theory.  In this context,
Anderson\cite{anderson,PWA} claims that in 2-dimensions
there is a failure of fermi
liquid theory for a dilute gas of fermions even with a weak short-range
two body repulsion.
Anderson bases his arguments on two important facts: a) the
presence of a two particle anti-bound state for arbitrarily small
repulsion in a
Hubbard model in 2-dimensions\cite{gbth},
and b) presence of a finite two particle phase
shift in the forward scattering channel on the fermi surface.
Several
authors\cite{renderia,RDS,fuku,stamp,fabri,wilkins,mattis,khlebnikov}
have studied this problem, and most of them have
concluded \cite{renderia,RDS,fuku,stamp,fabri}
that at the level of perturbation theory there is no indication of
a failure of fermi liquid theory.
Anderson\cite{PWA} has also
questioned the appropriateness of certain aspects of
conventional perturbation theory
to settle this subtle issue.

In this letter we study a dilute fermi gas having a weak short-range two
particle
repulsion using  simple perturbation theory.  In particular we look
at the energy correction to a pair of fermions with
arbitrary momenta $\hbar {\bf k}_1$ and
$\hbar {\bf k}_2$ inside the fermi sea and infer the presence of
an induced
${1\over{r^2}}$ interaction among the constituent fermions.

We identify two terms $\Sigma_{{\bf k}_1,{\bf k}_2}$ and
$\Sigma_{{\bf k}_2,{\bf k}_1}$, as kinetic
energy shifts of a
fermion at ${\bf k}_1$ due to the presence of another
at ${\bf k}_2$ and vice versa.
Their sum $\Sigma_{{\bf k}_1,{\bf k}_2}
+ \Sigma_{{\bf k}_2,{\bf k}_1}$ is given by the cross channel term,
in the language of diagrams
\cite{stamp}.  In Landau's fermi liquid theory, this
sum is the cross channel contribution to the Landau parameter for the
quasi particles.   An important point is that
$\Sigma_{{\bf k}_1,{\bf k}_2}$ and $\Sigma_{{\bf k}_2,{\bf k}_1}$ are
singular functions of $\mid {\bf k}_1 - {\bf k}_2 \mid$.

According to us, the natural decomposition of the cross channel diagram
into two terms, $\Sigma_{{\bf k}_1,{\bf k}_2}$
and $\Sigma_{{\bf k}_2,{\bf k}_1}$, and their singular character
suggests the following:
i) the terms $\Sigma_{{\bf k}_1,{\bf k}_2}$ and
$\Sigma_{{\bf k}_2,{\bf k}_1}$ are physically meaningful separately and
ii) we should look at the meaning and consequences of their
singularities
before we
proceed to consider the sum $\Sigma_{{\bf k}_1,{\bf k}_2} +
\Sigma_{{\bf k}_2,{\bf k}_1}$ as a Landau parameter
for the fermion quasi particles.  Our suggestion becomes even more
meaningful if the source of the singularity is capable
of destroying the
fermi liquid ground state.

Having emphasized the importance of the singular character
of $\Sigma_{{\bf k}_1,{\bf k}_2}$ and $\Sigma_{{\bf k}_2,{\bf k}_1}$,
we proceed and find that it can be caused by an
induced ${1\over{r^2}}$ two body
interaction among the constituent fermions\cite{note1}.
This induced two particle interaction is interpreted to arise from {\em
the
elimination of virtual scattering to high energy states in the presence
of the fermi sea, that is included in the perturbation
theory}\cite{note2}.

In the spirit of renormalised perturbation theory, the induced 2-body
potential is used in our next step as the effective interaction between
the fermions.  In 2-d, the long range two body potential
${1\over{r^2}}$ causes
a finite phase shift in the s-channel as the relative momentum
tends to zero.  This finite phase shift implies a scattering
length `a' which diverges as the size of the system.  Hence the
conventional fermi liquid perturbation expansion in terms of $k_F a$
fails, indicating an instability of the fermi liquid ground state.
Here $k_F$ is the fermi wave vector.

At the end of the letter we discuss how our work differs from
ealier works\cite{note3}, in particular that of Stamp\cite{stamp}.
Important
differences of our work from that of Anderson are also pointed
out at the end.

We now proceed to calculate the correction to the ground state energy
following the method of Abrikosov and Khalatnikov\cite{abri} which was
used recently by some authors for the 2-d case\cite{RDS,stamp}.
This method
is essentially an expansion in powers of the scattering length (a
physically measurable  quantity)
rather than the strength of the bare short range interaction.  We
consider the following Hamiltonian:
$$ H = \sum {{\hbar^2 k^2}\over 2m_e} c_{{\bf k}\sigma}^\dagger
 c_{{\bf k}\sigma} + {1 \over L^2}
 \sum U_q c^\dagger_{{\bf k - q}\sigma}
 c_{{\bf k}\sigma} c^\dagger_{{\bf k' + q}-\sigma}c_{{\bf k'},
 -\sigma}$$
where $c$'s are the fermion operators and $U_q$ is the  Fourier
transform of a short range two body interaction $ U(r)$.  We
can also view the above as the low density limit of the Hubbard
model in two dimensions, when $U_q$ has no q dependence. Following
Stamp and collaborators \cite{stamp} we define a dimensionless
interaction constant $\alpha$.   To second
order in $\alpha$, the ground state energy shift\cite{stamp,abri} is
given by
$$\sum {{\hbar^2 k^2}\over 2m_e} n_{{\bf k} \sigma }
 + {\hbar^2 \alpha \over m_e L^2}\sum n_{{\bf k}_1
 \uparrow}n_{{\bf k}_2 \downarrow}
  - \left ({\hbar^2 \alpha \over m_e L^2}\right )^2 \times$$
 $$\times \sum {n_{{{\bf k}_1{\sigma}_1}} n_{{{\bf k}_2{\sigma}_2}}
 [(1 - n_{{{\bf k}_3{\sigma}_3}})(1 - n_{{{\bf k}_4{\sigma}_4}})
 - 1]\over {\epsilon_{{\bf k}_1} + \epsilon_{{\bf k}_2} -
 \epsilon_{{\bf k}_3} - \epsilon_{{\bf k}_4} }}
 \times$$
\begin{equation}
\times\delta_{\sigma_1 -\sigma_2}
\delta_{\sigma_3 -\sigma_3}
  {\bf \delta}( {\bf k}_1 + {\bf k}_2 - {\bf k}_3 -{\bf k}_4 ).
\end{equation}
The parameter $\alpha$ is defined in terms of
$U_o$ as
\begin{equation}
\alpha = {m_e U_o \over \hbar^2}  +
{U_o^2 \over L^2} \sum {\delta ({\bf k}_1 +
{\bf k}_2 - {\bf k}_3 - {\bf k}_4 ) \over {\epsilon_{{\bf k}_1} +
\epsilon_{{\bf k}_2} - \epsilon_{{\bf k}_3} - \epsilon_{{\bf k}_4} }}.
\end{equation}
The dependence of $\alpha$ on ${\bf k}_1$ and ${\bf k}_2$ can be
ignored for low
densities.

The first term in equation (1) is the unperturbed
kinetic energy in the ground state.  The second term is the
Hartree term.  The third term is the most interesting term for us.
Let us consider two electrons in states
${{\bf k}_1{\sigma}_1}$ and ${{\bf k}_2{\sigma}_2}$ .  They
contribute an energy $ {\hbar^2
\over {2m_e}}(k^2_1 + k^2_2) $ to the unperturbed ground state
energy eigenvalue.  We would like to find out the correction
to this two particle contribution from the many body processes
considered within our perturbation theory, which is a
one step renormalisation.  This correction is given by
\begin{equation}
 n_{{\bf k}_1{\sigma}_1} n_{{\bf k}_2{\sigma}_2}
 \left ( \Delta E_{{\bf k}_1 , {\bf k}_2 }\delta_{\sigma_1,-\sigma_2}
 + \Sigma_{{\bf k}_1 , {\bf k}_2 } +
 \Sigma_{{\bf k}_2 , {\bf k}_1} \right ).
\end{equation}
Here the first term $\Delta E({\bf k}_1 , {\bf k}_2 )$ is a
symmetric function of
${\bf k}_1$ and ${\bf k}_2$ and is the Cooper channel contribution.
This is not of
interest to us, as it does not lead to any momentum shift,whose
value is comparable to the k-space lattice
spacing $\pi \over L$. Hence we will
not consider this any more. The second and third terms are the cross
channel terms that are unusual\cite{note4}.  Their existence
solely depends on the fact that we have
identical particles and a fermi sea.  In the absence of the fermi sea
they are simply absent in the second order energy correction.
Also, $\Sigma_{{\bf k}_1 , {\bf k}_2 }$ is not a
symmetric function of ${\bf k}_1$ and ${\bf k}_2$:
\begin{equation}
\Sigma_{{\bf k}_1,{\bf k}_2} \equiv
     \left ( {\hbar^2 \alpha} \over{m_e L^2}
    \right )^2
   \sum_{{\bf k}_3 ,{\bf k}_4 } {{ n_{{{\bf k}_3{\sigma}_3}}
  {\bf \delta}( {\bf k}_1 - {\bf k}_2 + {\bf k}_3 - {\bf k}_4 )}
  \over{\epsilon_{{\bf k}_1} - \epsilon_{{\bf k}_2} +
  \epsilon_{{\bf k}_3} - \epsilon_{{\bf k}_4}}}.
\end{equation}
Since we are interested in finding how occupied states within
the fermi sea are affected, we will only consider the
case  $k_1, k_2 < k_F$.
The summation over ${\bf k}_3$ and ${\bf
k}_4$ is easily simplified to give the following
two dimensional integral:
\begin{equation}
{1\over {\mid {\bf k}_1 - {\bf k}_2 \mid}}
\int^{2\pi}_{0} \int^{k_F}_0 \frac{k~dk~d\theta}{{\bf
k}_1 . {\hat {\bf k}}_{12} - k cos\theta},
\end{equation}
where ${\hat k}_{12}$ is the unit vector along the
direction of ${\bf k}_1 - {\bf k}_2$. This
integral can be  performed\cite{stamp} to give us the
result
\begin{equation}
      \Sigma_{{\bf k}_1,{\bf k}_2} =  \frac{\hbar^2 \alpha^2}
      {2\pi m_e L^2} \frac{{\bf k}_1.({\bf k}_1 -
      {\bf k}_2)}{({\bf k}_1 -{\bf k}_2 )^2} .
\end{equation}
for $\mid {\bf k}_1 \mid, \mid {\bf k}_2 \mid < k_F$.

We give a new interpretation to this singular term.
The above expression has the form of an energy increase arising
from a Galilean boost - it represents the kinetic energy change
arising from a momentum shift.  The fermion at ${\bf k}_1$
experiences a momentum shift owing to the presence of a fermion
at ${\bf k}_2$.  The momentum shift is
\begin{equation}
      \delta {\bf p}_{1,2} =  \frac{\hbar^2 \alpha^2}{2\pi L^2}
      \frac{{\bf k}_1.({\bf k}_1 -
      {\bf k}_2)}{({\bf k}_1 -{\bf k}_2 )^2} .
\end{equation}
If we consider two adjacent points in k-space, i.e.,
${\bf k}_1 - {\bf k}_2
= {\pi \over L}$, the above momentum shift is
$$
\delta p_{1,2} =
       \frac{\hbar^2 \alpha^2}{2\pi^2 L},
$$
which is of the order of the spacing in momentum space and
hence indicates a finite phase shift
$$
\delta_{ph} \approx \frac{\alpha}{2 \pi^2}.
$$
This expression agrees with Anderson's phase shift
calculation\cite{anderson}
for small U.
It is interesting to note that the
energy shift (equation 6) has the same form
that Anderson\cite{anderson} proposed as a
singular forward scattering term.  However, we would like to
re-emphasize that ours is not an expression for a
scattering amplitude but
a correction to the kinetic energy of a particle in scattering
state ${\bf k}_1$ due to the existence of another particle in
scattering state ${\bf k}_2$.  We will now show that
this singular form of kinetic energy shift arises if there is a longe
range potential of the form $1\over {r^2}$ in
2-dimension.

In what follows we show that a two body problem with a repulsive
potential of ${1\over{r^2}}$ in two dimensions  produces the same
analytic form of
momentum shift as given by equation (5)
{}From this we shall infer the
presence of an induced ${1\over{r^2}}$ potential between two particles
at long distances in our many body problem.

In a two body problem {\em if we know the
momentum shift of scattering states for all ${\bf k}$,  from the
form of the kinetic energy correction, we can
infer the asymptotic form of the effective two body potential}.
This is
because the momentum shift directly represents the
modification of the wave function in the asymptotic region,
which in turn is determined by the phase shift due to the two
body scattering.  Finally, once we know the two body
phase shift for two arbitrary scattering states, we can find the
asymptotic behaviour of the potential.

We concentrate on
plane wave states and find how their
energies get
modified due to any phase shifts in the various angular
momentum channels.  We will consider the repulsive potential
$V(r) = {\lambda\over {r^2}}$
between two particles and consider this in the
relative co-ordinate system.  Here $r$ is the relative separation
between the two particles and $\lambda$ is a constant.  We will assume
the following boundary condition on a circle of radius R  about the
origin:
$\psi (r) = 0 ~~~ {\rm for} ~~ r = R$.
If we consider a two particle problem in a finite domain like
a disc, the boundary condition in terms of the relative co-ordinates
is not simple because of the coupling of the center of mass and
relative co-ordinates.  However, the
results are not qualitatively modified by our simplified boundary
condition.
In the relative co-ordinate system,
the Schrodinger equation becomes simple and the scattering
states are characterised by radial and angular
momentum quantum numbers
$q$ and $m$.
The scattering states are
Bessel functions, that have the asymptotic form:
\begin{equation}
J_{\sqrt{ m^2 + \lambda_o}} (qr) \approx {\sqrt{2\over \pi qr}}
cos \left ( qr - {\pi \over 2} \sqrt{m^2 + \lambda_o}  - {{\pi}\over
4}\right )
\end{equation}
Using this asymptotic form  and imposing our boundary condition
it is easy to see that the  phase shift of the $m$-th partial
wave is
\begin{equation}
\delta_m \sim {{\lambda_o\pi} \over{4 m}}
\end{equation}
where $ \lambda_o = {2m_e \over \hbar^2} \lambda $.
The above states are radial eigen functions.  However, we are
interested in seeing how plane wave states get modified in the presence
of interaction.  We can easily obtain the phase shift suffered by the
scattering states that are plane waves  by the following wave
packet analysis\cite{anderson}.  Recall that partial waves are
obtained by
coherent superposition of all plane waves having the same
magnitude of the wave vector but with various directions in k-space
with
appropriate phase factors.  In the same way we can reconstruct
plane waves from the partial waves.  In doing so, only the large
$m$ partial states contribute dominantly.  In fact, if $q$ is the
value of the radial momentum, the partial waves that contribute
dominantly have the value of $m$ given by
$     {2\pi m \over R  } \approx q $.  Substituting this value
of $m$ in equation (9) we get the phase shift suffered by the
plane wave:
$$
\delta_{ph}(q) \approx {\lambda_o \pi^2  \over {2qR}}
$$
This phase shift is finite when q takes the least value of ${\pi
\over L}$. That is, $\delta_{ph}(q = {\pi \over L}) = {{\lambda_0
\pi} \over {2R}}$.  This finite phase shift in the s-channel as
$q \rightarrow 0$ is  well known for the ${1\over{r^2}}$ potential in
2-dimensions.
{}From this we find that the momentum shift suffered by the plane
wave with wave vector ${\bf q}$ is given by
$$
\delta {\bf q} \approx {\lambda_o \pi^2  \over {2qR^2}}\hat{q}
$$
By symmetry, the direction of momentum shift is in the same
direction as ${\bf q}$.  Notice that the momentum shift has a
singular dependence on $q$.  However, the change in kinetic energy in
the  relative co-ordinate system is
$$
  -{ \hbar^2\over {2m_e}} {\lambda_o \pi^2 \over {2 R^2}}
  {\bf q}.{{\bf q}\over{\vert {\bf q} \vert^2}}
$$
which is non-singular and independent of $q$.
Thus the  ${1\over{r^2}}$ potential in 2-dimensions
is anomalous in the sense that for two particle
plane wave states, the
kinetic energy shift of relative motion
has no singular dependence on the relative
momentum $q$ even though momentum
shift is singular.
We can substitute ${\bf q} = {\bf k}_1 - {\bf k}_2$ to go to
the laboratory frame
and get the shift in the kinetic
energy change of  particle $1$
$$
  -{ \hbar^2\over {2m}} {\lambda_o \pi^2 \over {2 R^2}}
  {\bf k}_1 .{({\bf k}_1 - {\bf k}_2 )
  \over{ ({\bf k}_1 - {\bf k}_2)^2}}
$$
and similarly for particle $2$.   This energy shift has the same
form as
the energy shift that we obtained by
perturbation theory for any two
occupied plane wave states inside the fermi sea (equation 6). Thus the
pseudo potential that acts between two electrons in the the occupied
states below the fermi surface is of ${1\over{r^2}}$ type at
long distances.
By comparing this energy shift
with equation (5) we find that the the strength of
the ${1\over{r^2}}$ term
is given by
$$
\lambda =  {\hbar^2 \alpha^2 \over m_e \pi^3}.
$$
Notice that our analysis only brings out the long range
part of the effective interaction.  The short range divergence
is cut off by the actual potential.

Having found a long range renormalised or effective interaction between
any two constituent particles in the ground state
using perturbation theory, one has to
use this as the starting interaction
in the spirit of renormalisation procedure to
find the properties of the final ground state.
We cannot use the conventional perturbation theory with the
${1\over{r^2}}$ potential.  This is because this repulsive potential
has a scattering length in 2-d which
diverges as $L$, the size of the
system.  This is to be contrasted with the scattering length for
the short range repulsive potential in 2-d, which diverges as
log$L$.  This logarithmic divergence is still manageable in
conventional perturbation theory.
Once we have a stronger divergence as $L$ there seems to be no
way of controlling the conventional perturbation expansion
thereby  indicating an instability of the fermi liquid ground
state.

To get around this difficulty, we have formulated\cite{GB} a new
approach, which enables us to write down the asymptotic behaviour
of low
energy many body wave functions that exhibit a non-fermi
liquid behaviour.  We will discuss this in a
forthcoming paper\cite{GB}.

What does our analysis predict for the known cases of 1 and 3-d
interacting fermi gas?   In 1-d we get  a strong signal from the
cross terms apart from other terms, indicating the failure of
fermi liquid theory.
In 3-d, the cross channel contribution
$\Sigma_{{\bf k}_1 ,{\bf k}_2 }$ has a
singular form
\begin{equation}
\Sigma_{{\bf k}_1,{\bf k}_2}\sim  \frac{1}{L^3 k_F}
\frac{{\bf k}_1 \cdot ({\bf k}_1 -{\bf k}_2)}
{({\bf k}_1 - {\bf k}_2)^2}.
\end{equation}
as ${\bf k}_1$ and ${\bf k}_2$ approaching each other.
However, the denominator has an $L^3$ instead of an $L^2$ as in two
dimensions.
We can read off the phase shift and momentum shift
when $ {\bf k}_1 \rightarrow {\bf k}_2$.
The phase shift is $\delta_{ph} \approx {1 \over L}$ , which
vanishes as $L \rightarrow \infty$. The momentum shift vanishes
as ${1\over L^2}$. Therefore, fermi liquid theory survives in 3-d
for small repulsive interaction.  Thus our results are
consistent with known results in one and two dimensions.

In conclusion, we would like to make a comparision of our work
with that of Stamp\cite{stamp}.  Stamp followed
Landau's theory, like other authors. In
addition he was
the first to notice the two singular pieces of the cross
channel contribution.  However, he attributed meaning only to
the sum $\Sigma_{{\bf k}_1,{\bf k}_2} +
\Sigma_{{\bf k}_2,{\bf k}_1}$ (which is non-singular) as a cross
channel contribution to the Landau parameter.  This led him to
conclude that at that level fermi liquid state is stable.
On the other hand, we point out that this can be done
only as long as there are no  parts of the energy correction
that is singular, which in principle could indicate the
presence of a finite phase shift as the relative momentum tends
to zero.
For example, in three dimension there
is no such correction and fermi liquid theory survives in the sense
that there is no finite phase shift.  Once a term
indicating the presence of a finite phase shift is present, it
signals an instability of the fermi liquid state and we have to
find the induced interaction that is responsible for the
momentum shift and then proceed to get the
ground state in the presence of this induced interaction.
There are other cases in 2-dimensions, where fermi liquid theory seems
to fail\cite{wen}.

Even though our approach is inspired by Anderson's works,
it has the following differences:
i) we are not calculating the phase shift in the
sense what Anderson does
ii)The anomalous energy shift has a singular form
very much like the singular forward scattering that Anderson
proposed - however, what we are
calculating is neither a Landau parameter nor a scattering amplitude,
but simply an energy shift.
iii) our kinetic energy shift which signals the
presence of an induced ${1\over{r^2}}$ potential occurs for
all pairs of electrons inside the fermi sea - it is not
confined to states close to the fermi surface.  In a sense we
give a first microscopic derivation for the induced long range
interaction that Anderson has conjectured.

I thank Hide Fukuyama for sending me Stamp's
paper\cite{stamp}.  My recent visits sharpened some of my
thinking.  I thank S. Doniach, R.B. Laughlin (Stanford),
P.W. Anderson, D.
Sherrington (Oxford), P. Guinea and D. Campbell (San Sebastian)
for hospitality.  I thank Muthukumar for several proof
readings.


\begin{references}
\bibitem[\ddag]{email}
baskaran@imsc.ernet.in
\bibitem{anderson} P. W. Anderson, Chapter 6 of Princeton RVB book(to
be published);
P. W. Anderson and Y. Ren, High Temperature
Superconductors, Editors K. S. Bedell et al, Proc. of the International
Conference on `` Physics of the Highly Correlated Electron Systems " ,
Los Alamos 1989 ( Addison-Wesley, NY 1989 )pp 1-33.
Physica Scripta, {\bf T42},11 (1992);
Progress of Theoretical Physics, Supplement {\bf 107},41,(1992);
Phys. Rev. Letters {\bf 67},2092(1991);
Phys. Rev. Letters {\bf 67},384(1991);
Phys. Rev. Letters,{\bf 64},2306 (1990); and
Phys. Rev. Letters,{\bf  66},3226 (1991).
\bibitem{PWA} P. W. Anderson, Phys. Rev. Letters,{\bf71},1220
(1993).
\bibitem{gbth} G. Baskaran and T. Hsu, Unpublished, 1987
\bibitem{renderia} J. Engelbrecht and M. Randeria,
Phys. Rev. Letters, {\bf 65},1032
(1990) and {\bf 66},325 (1991).
\bibitem{RDS}J.R. Engelbrecht et.al., Phys. Rev. {\bf 45},10135
(92).
\bibitem{fuku} H. Fukuyama et.al, J. Phys. Soc. Jpn,
{\bf 60}, 372(1991); S. Yarlagadda and S. Kurihara,
Phys. Rev. {\bf 48}, 10567 (1993).
\bibitem{stamp}
 P. C. E. Stamp, J. de Physique I,{\bf 3},625 (1993); and references
 therein .
\bibitem{fabri} G. Benfatto and G. Gallavotti, J. Stat. Phys. {\bf 59},
541 (1990); Phys. Rev. {\bf 42}, 9967 (1990);
R. Shankar, Int. J. Mod. Phys. {\bf B6},749 (1992);
Yale preprint (1993); C. Castellani et. al., Phys. Rev. Lett.,
{\bf 72},316 (1994);M. Fabrizio et.al, Phys. Rev. {\bf B44},1033
(1991).
\bibitem{wilkins} C. Hodges et.al., Phys. Rev.
{\bf B4}, 302(1971); P. Bloom, Phys. Rev. {\bf B12}, 125 (1975) ;
M. B. Vetrovec, G. M.
Caneiro, Phys. Rev. {\bf B22}, 1250 (1980).
\bibitem{mattis} Hua Chen and Daniel Mattis, Utah preprint
(1993)
\bibitem{khlebnikov}S.Yu. Khlebnikov, UCLA preprint 91/TEP/50
and 92/TEP/10;  R. Valenti and C. Gros, Phys. Rev. Lett.,
{\bf 68}, 2412 (1992).
\bibitem{abri} A. A. Abrikosov, I. Khalatnikov, JETP {\bf
6}, 888(1958)
; Landau and Lifshitz series, Statistical Physics Part 2, Vol
.9, (Pergamon Press, NY 1980) sec. 6.
\bibitem{note1}In the process we also show that the energy
shifts of two particle scattering states for ${1\over{r^2}}$
potential in
2-dimension is strange in the sense that the relative momentum
shift is a singular function of $({\bf k}_1 - {\bf k}_2)$, whereas the
kinetic energy shift of relative motion is not.
\bibitem{note2}That is, this
perturbation theory is viewed as a one step renormalisation procedure,
in the same sense as that of a degenerate perturbation theory
that one does
for large U Hubbard model, which at half filling
induces a pairwise Heisenberg spin coupling.
\bibitem{note3}Our approach should be contrasted with the
other approaches,
where one attempts to get the two particle scattering amplitude
using a perturbation theory for electron or hole self energy or
the two particle T matrix.  We wish to clarify that we do
not go off-shell and
look at quasi particles or quasi holes.
\bibitem{note4}These cross channel terms are anomalous -
they represent certain
anomalous processes.
The term $\Sigma_{{\bf k}_1 ,{\bf k}_2 }$ is the second
order energy correction of an interacting  `particle'
and a `hole': particle in state ${{\bf k}_1{\sigma}_1}$ and a hole in
state $ - {{\bf k}_2{\sigma}_2}$ with kinetic
energy $-\epsilon_{{\bf k}_2}$.  Notice that both
the particle and hole have momenta lying inside the fermi sea !
In this sense they are anomalous and represent processes that seem
to violate
the Pauli principle and the fundamental definition of particle
and hole.  It is easily seen that as the fermi momentum vanishes
this term also vanishes.
\bibitem{GB} G. Baskaran, unpublished manuscript
\bibitem{wen}P.A. Bares and X.G. Wen, Phys. Rev. {\bf B48},
8636 (1993);J. Polchinski, NSF-ITP-93-33;
C. Nayak and F. Wilczek, PUPT 1438,IASSNS 93/89.

\end{references}
\end{document}